\documentclass{article}

\usepackage{laf}
\usepackage{epsfig}
\usepackage{rotating}

\newcommand{\la}{\langle}
\newcommand{\ra}{\rangle}
\newcommand{\be}{\begin{eqnarray}}
\newcommand{\ee}{\end{eqnarray}}

\begin{document}

\bibliographystyle{unsrt}

\title{Complete quantum teleportation by nuclear magnetic resonance}

\author{M.~A.~Nielsen$^{*\dag}$, E.~Knill$^{\ddag}$, \&
R.~Laflamme$^*$ \vspace{0.2cm}  \\
$*$ Theoretical Astrophysics T-6, \\ 
MS B-288,
Los Alamos National Laboratory, \\
Los Alamos, NM 87545 \vspace{0.2cm} \\ 
$\dag$Department of Physics and Astronomy, \\
University of New Mexico, \\
Albuquerque, NM 87131-1131 \vspace{0.2cm} \\
$\ddag$Computer Research and Applications CIC-3, \\
MS B-265,
Los Alamos National Laboratory, \\
Los Alamos, NM 87545}

\maketitle

\textbf{Quantum mechanics provides spectacular new information
processing abilities \cite{Bennett95a,Preskill98a}.  One of the most
unexpected is a procedure called \textsl{ quantum teleportation}
\cite{Bennett93a} that allows the quantum state of a system to be
transported from one location to another, without moving through the
intervening space. Partial implementations of teleportation
\cite{Bouwmeester97a,Boschi98a} over macroscopic distances have been
achieved using optical systems, but omit the final stage of the
teleportation procedure. Here we report an experimental implementation
of the full quantum teleportation operation over inter-atomic
distances using liquid state nuclear magnetic resonance (NMR).  The
inclusion of the final stage enables for the first time a
teleportation implementation which may be used as a
\textsl{subroutine} in larger quantum computations, or for quantum
communication. Our experiment also demonstrates the use of
\textsl{quantum process tomography}, a procedure to completely
characterize the dynamics of a quantum system.  Finally, we
demonstrate a controlled exploitation of decoherence as a tool to
assist in the performance of an experiment.  }


In classical physics, an object can be teleported, in principle, by
performing a measurement to completely characterize the properties of
the object. That information can then be sent to another location, and
the object reconstructed.  Does this provide a complete reconstruction
of the original object?  No: all physical systems are ultimately
quantum mechanical, and quantum mechanics tells us that it is
impossible to completely determine the state of an unknown quantum
system, making it impossible to use the classical measurement
procedure to move a quantum system from one location to another.

Bennett {\em et al} \cite{Bennett93a} have suggested a remarkable
procedure for teleporting quantum states.  Quantum teleportation may
be described abstractly in terms of two parties, Alice and Bob.  Alice
has in her possession an {\em unknown} state $|\Psi\ra=\alpha|0\ra
+\beta|1\ra$ of a single quantum bit (qubit) -- a two level quantum
system. The goal of teleportation is to transport the state of that
qubit to Bob. In addition, Alice and Bob each possess one qubit of a
two qubit entangled state, \be |\Psi\ra_A \left(|0\ra_A|0\ra_B
+|1\ra_A|1\ra_B \right), \ee where subscripts $A$ are used to denote
Alice's systems, and subscripts $B$ to denote Bob's system. Here and
throughout we omit overall normalization factors from our equations.

This state can be rewritten in the {\em Bell basis} ($\left(|00\ra \pm
|11\ra \right), \left( |01\ra \pm |10\ra \right)$ for the first two
qubits and a conditional unitary transformation of the state
$|\Psi\ra$ for the last one, that is, \be & &
\left(|00\ra+|11\ra\right)|\Psi\ra + \left(|00\ra-|11\ra\right)
\sigma_z |\Psi\ra + \nonumber \\ & & \left(
|01\ra+|10\ra \right) \sigma_x |\Psi\ra + \left( |01\ra - |10\ra
\right) \left(-i \sigma_y |\Psi\ra \right), \ee where
$\sigma_x,\sigma_y,\sigma_z$ are the Pauli sigma operators
\cite{Sakurai95a}, in the $|0\ra, |1\ra$ basis. A measurement is
performed on Alice's qubits in the Bell basis. 
Conditional on these measurement outcomes it is easy
to verify from the previous equation that Bob's respective states are
\be |\Psi\ra; \,\,\,\, \sigma_z |\Psi\ra; \,\,\,\, \sigma_x |\Psi\ra;
\,\,\,\, -i \sigma_y |\Psi\ra.  \ee Alice sends the outcome of her
measurement to Bob, who can then recover the original state $|\Psi\ra$
by applying the appropriate unitary transformation $I, \sigma_x,
\sigma_z$, or $i \sigma_y$, conditional on Alice's measurement
outcome. Notice that the quantum state transmission has not been
accomplished faster than light because Bob must wait for Alice's
measurement result to arrive before he can recover the quantum state.

Recent demonstrations of quantum teleportation
\cite{Bouwmeester97a,Boschi98a} omitted the final stage of
teleportation, the unitary operators applied by Bob conditional on the
result of Alice's measurement. This prevents complete recovery of the
original state.  Instead, the earlier experiments relied on classical
post-processing of the data after completion of the experiment to
check that the results were consistent with what one would expect {\em
if the conditional operations had, in fact, been performed}.  Our
experiment implements the full teleportation operation.  The most
important implication of the inclusion of this difficult extra stage
is that our teleportation procedure can, in principle, be used as a
{\em subroutine} in the performance of other quantum information
processing tasks.  Teleportation as a subroutine is important in
potential applications to quantum computation and communication
\cite{Bennett96a,Cirac97a}, although in the present system, moving the
C2 qubit to H may be accomplished more efficiently by techniques other
than teleportation.


Our implementation of teleportation is performed using liquid state
nuclear magnetic resonance (NMR), applied to an ensemble of molecules
of labeled trichloroethylene (TCE).  The structure of the TCE molecule
may be depicted as
\begin{eqnarray}
\unitlength 0.4cm
\begin{picture}(8,2)(-1,1)
\put(2.9,2.10){\line(1,0){1.2}}
\put(2.9,1.90){\line(1,0){1.2}}
\put(1.9,1.7){Cl}
\put(4.3,1.7){C2}
\put(1.8,2.1){\line(-3,2){1.0}}
\put(-0.1,3.0){$\mathrm{C}\mathit{l}$}
\put(5.5,2.1){\line(3,2){1.0}}
\put(6.85,3.0){$\mathrm{C}\mathit{l}$}
\put(5.5,1.9){\line(3,-2){1.0}}
\put(6.85,0.6){$\mathrm{C}\mathit{l}$}
\put(1.8,1.9){\line(-3,-2){1.0}}
\put(0.0,0.6){H}
\end{picture}
\end{eqnarray}
To perform teleportation we make use of the Hydrogen nucleus (H), and
the two Carbon 13 nuclei (C1 and C2), teleporting the state of C2 to
H.  Figure 1(a) illustrates the teleportation process we used.  The
circuit has three inputs, which we will refer to as the {\em data}
(C2), {\em ancilla} (C1), and {\em target} (H) qubits. The goal of the
circuit is to teleport the state of the data qubit so that it ends up
on the target qubit.  We are therefore only teleporting the qubit a
few angstroms, making this a demonstration of the method of
teleportation, rather than a practical means for transmitting qubits
over long distances.


State preparation is done in our experiment using the gradient-pulse
techniques described by Cory {\em et al} \cite{Cory97a}, and phase
cycling \cite{Ernst90a,Grant96a}.  The unitary operations performed
during teleportation may be implemented in a straightforward manner in
NMR, using non-selective rf pulses tuned to the Larmor frequencies of
the nuclear spins, and delays allowing entanglement to form through
the interaction of neighboring nuclei \cite{Cory97a,Gershenfeld97a}.
Other demonstrations of quantum information processing with three
qubits using NMR are described in \cite{Cory98b,Laflamme98a,Cory98a},
and with two qubits in \cite{Chuang98a,Jones98a,Chuang98c,Jones98b}.


An innovation in our experiment was the method used to implement the
Bell basis measurement.  In NMR, the measurement step allows us to
measure the expectation values of $\sigma_x$ and $\sigma_y$ for each
spin, averaged over the ensemble of molecules, rather than performing
a projective measurement in some basis.  For this reason, we must
modify the projective measurement step in the standard description of
teleportation, while preserving the remarkable teleportation effect.

We use a procedure inspired by Brassard {\em et al}
\cite{Brassard98a}, who suggest a two-part procedure for performing
the Bell basis measurement. Part one of the procedure is to rotate
from the Bell basis into the computational basis, $|00\ra,|01\ra,
|10\ra,|11\ra$.  We implement this step in NMR by using the natural
spin-spin coupling between the Carbon nuclei, and rf
pulses.  Part two of the procedure is to perform a projective
measurement in the computational basis.  As Brassard {\em et al} point
out, the effect of this two part procedure is equivalent to performing
the Bell basis measurement, and leaving the data and ancilla qubits in
one of the four states, $|00\ra, |01\ra, |10\ra, |11\ra$,
corresponding to the different measurement results.

We cannot directly implement the second step in NMR.  Instead, we
exploit the natural phase decoherence occurring on the Carbon nuclei
to achieve the same effect.  Recall that phase decoherence completely
randomizes the phase information in these nuclei and thus will destroy
coherence between the elements of the above basis.  Its effect on the
state of the Carbon nuclei is to diagonalize the state in the
computational basis, \be \rho & \longrightarrow & |00\ra \la 00| \rho
|00\ra \la 00| + |01\ra \la 01| \rho |01\ra \la 01| +|10\ra \la 10|
\rho |10\ra \la 10| \nonumber \\ & & + |11\ra \la 11| \rho |11\ra \la
11|. \ee As emphasized by Zurek \cite{Zurek91a}, the decoherence
process is indistinguishable from a measurement in the computational
basis for the Carbons accomplished by the environment.  We do not
observe the result of this measurement explicitly, however the state
of the nuclei selected by the decoherence process contains the
measurement result, and therefore we can do the final transformation
conditional on the particular state the environment has selected.  As
in the scheme of Brassard {\em et al}, the final state of the Carbon
nuclei is one of the four states, $|00\ra,|01\ra,|10\ra,|11\ra$,
corresponding to the four possible results of the measurement.

Our experiment exploits the natural decoherence properties of the TCE
molecule.  The phase decoherence times ($T_2$) for C1 and C2 are
approximately $0.4s$ and $0.3s$.  All other $T_2$ and $T_1$ times for
all three nuclei are much longer, with a $T_2$ time for the Hydrogen
of approximately $3s$, and relaxation times ($T_1$) of approximately
$20-30s$ for the Carbons, and $5s$ for the Hydrogen.

Thus, for delays on the order of $1s$, we can approximate the total
evolution by exact phase decoherence on the Carbon nuclei. The total
scheme therefore implements a measurement in the Bell basis, with the
result of the measurement stored as classical data on the Carbon
nuclei following the measurement.  We can thus teleport the state from
the Carbon to the Hydrogen and verify that the state in the final
state decays at the Hydrogen rate and not the Carbon rate.

Re-examining figure 1(a) we see how remarkable teleportation is from
this point of view.  During the stage labeled ``Measure in the Bell
basis'' in figure 1(a), we allow the C1 and C2 nuclei to decohere and
thus be measured by the environment, destroying all phase information
on the data and ancilla qubits.  Experimentally, a standard NMR
technique known as {\em refocusing} employs rf pulses to ensure that
the data qubit effectively does not interact with the target
qubit. Classical intuition therefore tells us that the phase
information about the input state, $|\Psi\rangle$, has been lost
forever. Nevertheless, quantum mechanics predicts that we are still
able to recover the complete system after this decoherence step, by
quantum teleportation.


We implemented this scheme in TCE using a Bruker DRX-500 NMR
spectrometer.  Experimentally, we determined the Larmor and coupling
frequencies for the Hydrogen, C1 and C2 to be: \be \omega_H \approx
500.133491 \mbox{MHz}; & \,\,\,\, \omega_{C1} \approx 125.772580
\mbox{MHz}; &\,\,\,\, \omega_{C2} \approx \omega_{C1}- 911 \mbox{Hz}
\nonumber \\ & & \\ J_{H \, C1} \approx 201 \mbox{Hz}; & \,\,\,\,
J_{C1 \, C2} \approx 103 \mbox{Hz}. & \ee The coupling frequencies
between H and C2, as well as the Chlorines to H, C1 and C2, are much
lower, on the order of ten Hertz for the former, and less than a Hertz
for the latter.  Experimentally, these couplings are suppressed by
multiple refocusings, and will be ignored in the sequel.  Note that C1
and C2 have slightly different frequencies, due to the different
chemical environments of the two atoms.

We performed two separate sets of experiments. In one set, the full
teleportation process was executed, making use of a variety of
decoherence delays in place of the measurement. The readout was
performed on the Hydrogen nucleus, and a figure of merit -- the
entanglement fidelity -- was calculated for the teleportation process.
The entanglement fidelity is a quantity in the range $0$ to $1$ which
measures the combined strength of {\em all} noise processes occurring
during the process \cite{Schumacher96a,Barnum98a}. In particular, an
entanglement fidelity of $1$ indicates perfect teleportation, while an
entanglement fidelity of $0.25$ indicates total randomization of the
state.  Perfect {\em classical transmission} corresponds to an
entanglement fidelity of $0.5$ \cite{Schumacher96a,Barnum98a}, so
entanglement fidelities greater than $0.5$ indicates that
teleportation of some quantum information is taking place.

The second set of experiments was a control set, illustrated in
figure~1(b). In those experiments, only the state preparation and
initial entanglement of H and C1 were performed, followed by a delay
for decoherence on C1 and C2. The readout was performed in this
instance on C2, and the entanglement fidelity was calculated for the
process.


The results of our experiment are shown in figure~2, where the
entanglement fidelity is plotted against the decoherence delay. Errors
in our experiment arise from the strong coupling effect, imperfect
calibration of rf pulses, and rf field inhomogeneities. The estimated
uncertainties in the entanglement fidelities are less than $\pm 0.05$,
and are due primarily to rf field inhomogeneity and imperfect
calibration of rf pulses.

To determine the entanglement fidelities for the teleportation and
control experiments, we performed {\em quantum process tomography}
\cite{Poyatos97a,Chuang97a}, a procedure for obtaining a complete
description of the dynamics of a quantum system, as follows: The
linearity of quantum mechanics implies that the single qubit input and
output for the teleportation process are related by a linear quantum
operation \cite{Nielsen97c}.  By preparing a complete set of four
linearly independent initial states, and measuring the corresponding
states output from the experiment, we may completely characterize the
quantum process, enabling us to calculate the entanglement fidelity
for the process \cite{Chuang97a}.

Three elements ought to be noted in figure~2.  First, for small
decoherence delays, the entanglement fidelity for the teleportation
experiments significantly exceeds the value of $0.5$ for perfect
classical transmission of data, indicating successful teleportation of
quantum information from C2 to H, with reasonable fidelity.  Second,
the entanglement fidelity decays very quickly for the control
experiments as the delay is increased.  Theoretically, we expect this
to be the case, due to a $T2$ time for C2 of approximately $0.3s$.
Third, the decay of the entanglement fidelity for the teleportation
experiments occurs much more slowly.  Theoretically, we expect this
decay to be due mainly to the effect of phase decoherence and
relaxation on the {\em Hydrogen}.  Our experimental observations are
consistent with this prediction, and provide more support for the
claim that quantum data is being teleported in these experiments.

{\bf Acknowledgments} We thank David Cory, Chris Jarzynski, Jun Ye,
and Wojtek Zurek for useful discussions, the Stable Isotope Laboratory
at Los Alamos for use of their facility, and the National Security
Agency and Office of Naval Research for support.

Correspondence and requests for materials should be sent to
M.~A.~N. (mnielsen@theory.caltech.edu).

%
%

\onecolumn

{\Large \bf Figure Captions}

\vspace{2cm} Figure 1: Schematic circuits for (a) the quantum
teleportation experiment, and (b) the control experiment.  The
teleportation circuit is based upon that suggested by Brassard {\em et
al} $^{21}$.  Note that the control circuit simply omits two elements
of the teleportation experiment -- rotation from the Bell basis into
the computational basis, immediately before the decoherence step, and
the conditional unitary operation. Commented pulse sequences for our
experiment may be obtained on the world wide web $^{13}$.

\vspace{2cm} Figure 2: Entanglement fidelity (a measure of how well quantum
information is preserved) is plotted as a function of decoherence
time.  The bottom curve is a control run where the information remains
in C2.  The curve shows a decay time of approximately 0.5s.  The top
curve represents the fidelity of the quantum teleportation process.
The decay time is approximately 2.6s.  The information is preserved
for a longer time, corresponding approximately to the combined effects
of decoherence and relaxation for the Hydrogen, confirming the
prediction of teleportation.

\newpage

\begin{figure}[h]
\begin{centering}
\unitlength 0.8cm
\begin{picture}(15,7.3)(-2,-4)
\put(-1.3,3.2){(a)}
\put(-1.3,2.6){data}
\put(-1.29,2.25){(C2)}
\put(-1.45,1.6){ancilla}
\put(-1.26,1.25){(C1)}
\put(-1.4,0.6){target}
\put(-1.22,0.10){(H)}
\put(-0.2,2.45){$|\Psi\rangle$}
\put(0,1.5){$0$}
\put(0,0.4){$0$}
\put(-1.9,1.45){{\begin{sideways}\textbf{Alice}\end{sideways}}}
\put(-1.9,-0.05){{\begin{sideways}\textbf{Bob}\end{sideways}}}
\put(-2.0,1.10){\framebox(2.5,1.9){}}
\put(-2.0,-0.15){\framebox(2.5,1.1){}}
\put(0.5,2.5){\line(1,0){3}}
\put(0.5,1.5){\line(1,0){0.5}}
\put(0.5,0.5){\line(1,0){0.5}}
\put(1,0.2){\framebox(2,1.6){entangle}}
\put(3,1.5){\line(1,0){0.5}}
\put(3,0.5){\line(1,0){4.5}}
\put(3.5,1.2){\framebox(3.5,1.6){}}
\put(3.5,2.1){\makebox(3.5,0.5){measure in the}}
\put(3.5,1.4){\makebox(3.5,0.5){Bell basis}}
\put(7,2.5){\line(1,0){2.4}}
\put(7,1.5){\line(1,0){2.4}}
\put(7.5,0.0){\framebox(4,1){}}
\put(7.5,0.5){\makebox(4,0.5){conditional unitary}}
\put(7.5,0.0){\makebox(4,0.5){$U_{00},U_{01},U_{10},U_{11}$}}
\put(9.5,1){\line(0,1){0.4}}
\put(9.5,1.5){\circle{0.2}}
\put(9.5,1.6){\line(0,1){0.8}}
\put(9.5,2.5){\circle{0.2}}
\put(9.6,2.5){\line(1,0){2.4}}
\put(9.6,1.5){\line(1,0){2.4}}
\put(11.5,0.5){\line(1,0){0.5}}
\put(12.3,2.6){classical}
\put(12.5,2.3){data}
\put(12.3,1.6){classical}
\put(12.5,1.3){data}
\put(12.65,0.43){$| \Psi \rangle$}
%
%
\put(-1.3,-0.8){(b)}
\put(-1.3,-1.4){data}
\put(-1.29,2.25){(C2)}
\put(-1.45,-2.4){ancilla}
\put(-1.26,-2.75){(C1)}
\put(-1.4,-3.4){target}
\put(-1.22,-3.90){(H)}
\put(-0.2,-1.55){$|\Psi\rangle$}
\put(0,-2.5){$0$}
\put(0,-3.6){$0$}
\put(-1.9,-2.55){{\begin{sideways}\textbf{Alice}\end{sideways}}}
\put(-1.9,-4.05){{\begin{sideways}\textbf{Bob}\end{sideways}}}
\put(-2.0,-2.90){\framebox(2.5,1.9){}}
\put(-2.0,-4.15){\framebox(2.5,1.1){}}
\put(0.5,-1.5){\line(1,0){3}}
\put(0.5,-2.5){\line(1,0){0.5}}
\put(0.5,-3.5){\line(1,0){0.5}}
\put(1,-3.8){\framebox(2,1.6){entangle}}
\put(3,-2.5){\line(1,0){0.5}}
\put(3,-3.5){\line(1,0){5}}
\put(3.5,-2.8){\framebox(3.5,1.6){}}
\put(3.5,-2.25){\makebox(3.5,0.5){decohere}}
\put(7,-1.5){\line(1,0){1}}
\put(7,-2.5){\line(1,0){1}}
\put(8.3,-1.55){tomography}
\end{picture}
\end{centering}
\caption{ Schematic circuits for (a) the quantum
teleportation experiment, and (b) the control experiment.  The
teleportation circuit is based upon that suggested by Brassard {\em et
al} $^{21}$.  Note that the control circuit simply omits two elements
of the teleportation experiment -- rotation from the Bell basis into
the computational basis, immediately before the decoherence step, and
the conditional unitary operation. Commented pulse sequences for our
experiment may be obtained on the world wide web $^{13}$.}
\end{figure}
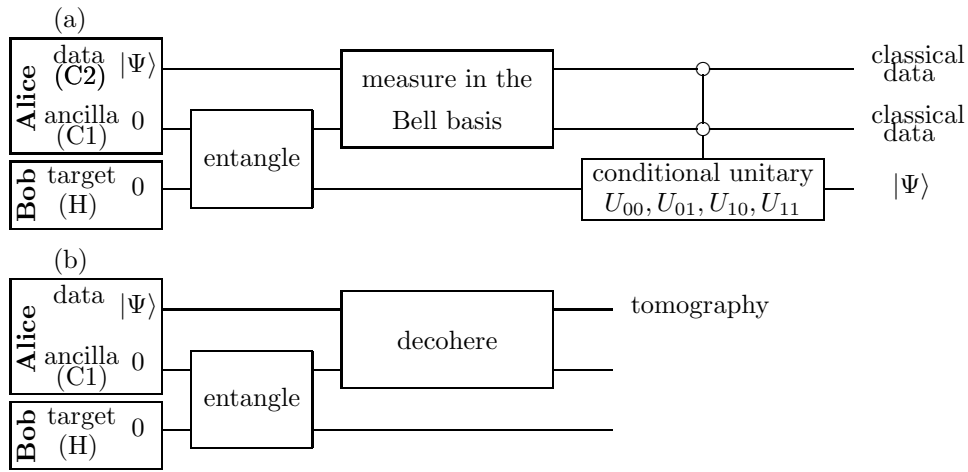


\newpage

\begin{figure}[t]
\epsfxsize 4.5in
\begin{center}
\epsfbox{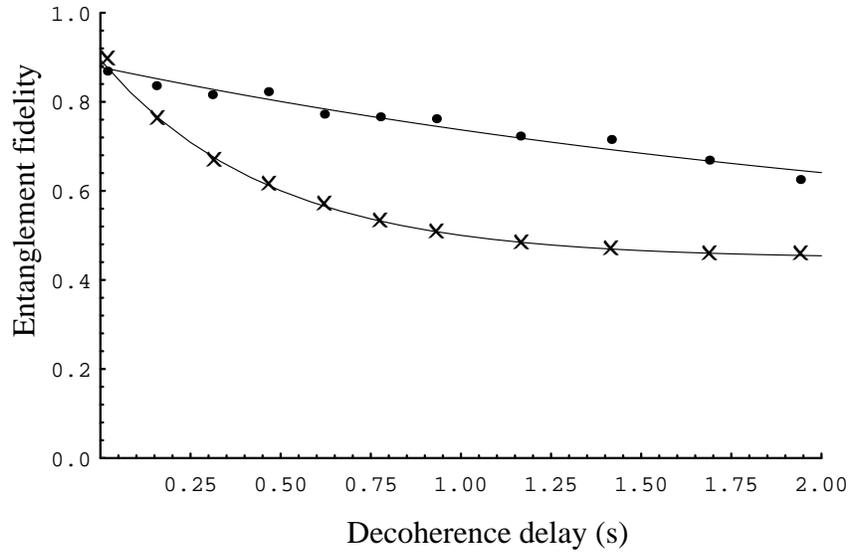}
\end{center}
\caption{Entanglement fidelity (a measure of how well quantum
information is preserved) is plotted as a function of decoherence
time.  The bottom curve is a control run where the information remains
in C2.  The curve shows a decay time of approximately 0.5s.  The top
curve represents the fidelity of the quantum teleportation process.
The decay time is approximately 2.6s.  The information is preserved
for a longer time, corresponding approximately to the combined effects
of decoherence and relaxation for the Hydrogen, confirming the
prediction of teleportation.}
\end{figure}

\end{document}